\input lanlmac

\def\k{\kappa}

\def\a{\alpha}
\def\b{\beta}

\def\g{\gamma}

\def\D{\Delta}

\def\p{\phi}

 \def \k {\kappa}

\def\l{\lambda}

\def\s{\sigma}
\def \sm {$\s$-model\ }

\def \ov {\over}
\def \ha { { 1\ov 2}}

 \def\ep {\epsilon}

\def \gp { g_5}
\def \F {{\cal F}}
\def\CF {{\cal F}}
\def \del { \partial}
\def \t {\theta}
\def \p {\phi}

\def\np {  Nucl. Phys. }
\def \pl { Phys. Lett. }

\def \pr  { Phys. Rev. }

\def \cD {{\cal D}}
\def \td {\tilde}

\def \log {{\rm log \ }}
\def \ln {{\rm \ ln \  }}

\def \ch {{\rm cosh \ }}

\def \l {\lambda}
\def \p {\phi}
\def \four{{\textstyle{1\over 4}}}

\gdef \jnl#1, #2, #3, 1#4#5#6{ {\sl #1~}{\bf #2} (1#4#5#6) #3}

\def \ch {\ {\rm cosh} \ }

\def \ln { {\rm ln } }

\def \cos { {\rm cos}  }

\def \l {\lambda}
\def \p {\phi}
\def \vp {\varphi}
\def  \g {\gamma}

\def\({\left (}
\def\){\right )}
\def\[{\left [}
\def\]{\right ]}

\def \pp {\phi}

\def \lr {\lref}

\lr \TT { C.~Lovelace,
``Stability Of String Vacua. 1. A New Picture Of
The Renormalization Group,''
Nucl.\ Phys.\ B {\bf 273}, 413 (1986).
B.~E.~Fridling and A.~Jevicki,
``Nonlinear Sigma Models As S Matrix Generating
Functionals Of Strings,''
Phys.\ Lett.\ B {\bf 174}, 75 (1986).
A. A.~Tseytlin,
``Vector Field Effective Action In The Open
Superstring Theory,''
Nucl.\ Phys.\  {\bf B276}, 391 (1986).
}

\lr\mec{A. A.~Tseytlin,
``Mobius Infinity Subtraction And Effective Action In Sigma Model
Approach To Closed String Theory,''
Phys.\ Lett.\  {\bf B208}, 221 (1988).
}

\lr\mecc{A. A.~Tseytlin,
``Conditions Of Weyl Invariance Of Two-Dimensional Sigma Model {}From
Equations Of Stationarity Of 'Central Charge' Action,''
Phys.\ Lett.\  {\bf B194}, 63 (1987).
}

\lr \callan{ Callan et al }

\lref\GH {G.~W.~Gibbons and S.~W.~Hawking,
``Action Integrals And Partition Functions In
Quantum Gravity,''
Phys.\ Rev.\ D {\bf 15}, 2752 (1977).
}

\lref \lovet {C. Lovelace, \jnl \np,  B273, 413, 1986;
B. Fridling and A. Jevicki, \jnl\pl, B174, 75, 1986.}
\lr\osb {H. Osborn, \jnl \np, B308, 629, 1988. }

\def\ha{{\textstyle{1\over2}}}

\lr \frts {E.S.  Fradkin  and A. A. Tseytlin, \jnl \pl, B158, 316, 1985;
\jnl \np, B261, 1, 1985.}
\lref \loveo {C. Lovelace, \jnl \pl,  B135, 75, 1984;
P. Candelas, G. Horowitz, A. Strominger and E. Witten,
\jnl\np, B258, 46, 1985.}
\lr \weh{C.M. Hull and P.K. Townsend, \jnl \np, B274, 349, 1986;
A. A. Tseytlin, \jnl \pl, B178, 34, 1986; \jnl \np, B294, 383, 1987.}

\lr \tsred{A.~A.~Tseytlin,
``Ambiguity In The Effective Action In String
Theories,''
Phys.\ Lett.\ B {\bf 176}, 92 (1986).
}

\lr \grwi {D.~J.~Gross and E.~Witten,
``Superstring Modifications Of Einstein's
Equations,''
Nucl.\ Phys.\ B {\bf 277}, 1 (1986).
}

\lr \polyakov { A. Polyakov, \jnl \pl, B103, 207, 1981.  }

\lr \FT { E.~S.~Fradkin and A.~A.~Tseytlin,
``Quantum String Theory Effective Action,''
Nucl.\ Phys.\ B {\bf 261}, 1 (1985).
 }

\lr\tspl {A.~A.~Tseytlin,
``On field redefinitions and exact solutions in
string theory,''
Phys.\ Lett.\ B {\bf 317}, 559 (1993)
[hep-th/9308042].
 }

\lr \panv { I.~Jack, D.~R.~Jones and J.~Panvel,
``Exact bosonic and supersymmetric string black
hole solutions,''
Nucl.\ Phys.\ B {\bf 393}, 95 (1993)
[hep-th/9201039].
 }

\lr\pert{ M.~J.~Perry and E.~Teo,
``Nonsingularity of the exact two-dimensional
string black hole,''
Phys.\ Rev.\ Lett.\ {\bf 70}, 2669 (1993)
[hep-th/9302037].
}

\lr \kkk { V.~Kazakov, I.~K.~Kostov and D.~Kutasov,
``A matrix model for the two dimensional black
hole,''
hep-th/0101011.
} 

\lr \FZZ  {V. Fateev, A. Zamolodchikov and Al. Zamolodchikov, unpublished.} 

\lref \dvv  { R.~Dijkgraaf, H.~Verlinde and
E.~Verlinde,
``String propagation in a black hole geometry,''
Nucl.\ Phys.\ B {\bf 371}, 269 (1992).
 }

\lr\rabi  {S.~Elitzur, A.~Forge and E.~Rabinovici,
``Some global aspects of string
compactifications,''
Nucl.\ Phys.\ B {\bf 359}, 581 (1991).
G.~Mandal, A.~M.~Sengupta and S.~R.~Wadia,
``Classical solutions of two-dimensional string
theory,''
Mod.\ Phys.\ Lett.\ A {\bf 6}, 1685 (1991).
}

 \lr \witt{ E.~Witten,
``On string theory and black holes,''
Phys.\ Rev.\ D {\bf 44}, 314 (1991).
}

\lr \gp { G.~W.~Gibbons and M.~J.~Perry,
``The Physics of 2-D stringy space-times,''
Int.\ J.\ Mod.\ Phys.\ D {\bf 1}, 335 (1992)
[hep-th/9204090].
}

\lr \ttt {A.~A.~Tseytlin,
``Exact solutions of closed string theory,''
Class.\ Quant.\ Grav.\ {\bf 12}, 2365 (1995)
[hep-th/9505052].
}

\def \t {\theta}
\def \M {{\cal M}}

\def \rf {\refs}

\lr \rrbh{
T.~Jacobson, G.~Kang and R.~C.~Myers,
``On black hole entropy,''
Phys.\ Rev.\ D {\bf 49}, 6587 (1994)
[gr-qc/9312023].
``Increase of black hole entropy in higher curvature gravity,''
Phys.\ Rev.\ D {\bf 52}, 3518 (1995)
[gr-qc/9503020].
G.~L.~Cardoso, B.~de Wit, J.~Kappeli and T.~Mohaupt,
``Supersymmetric black hole solutions with $R^2$
 interactions,''
hep-th/0003157.
S.~O.~Alexeev and M.~V.~Pomazanov,
``Black hole solutions with dilatonic hair in higher curvature gravity,''
Phys.\ Rev.\ D {\bf 55}, 2110 (1997)
[hep-th/9605106].
R.~C.~Myers,
``Black holes in higher curvature gravity,''
gr-qc/9811042.
}

\lr\nap{C.~R.~Nappi and A.~Pasquinucci,
``Thermodynamics of two-dimensional black holes,''
Mod.\ Phys.\ Lett.\ A {\bf 7}, 3337 (1992)
[gr-qc/9208002].
}

\lr\wald{ R.~M.~Wald,
``Black hole entropy is Noether charge,''
Phys.\ Rev.\ D {\bf 48}, 3427 (1993)
[gr-qc/9307038].
}

\lr\alpo{ 9605106 } 

\def \D {{\cal D}}
\def \w {{\rm w}}

\lr \bsfet {I.~Bars and K.~Sfetsos,
``Conformally exact metric and dilaton in string
theory on curved space-time,''
Phys.\ Rev.\ D {\bf 46}, 4510 (1992)
[hep-th/9206006].
``Exact effective action and space-time geometry n
gauged WZW models,''
Phys.\ Rev.\ D {\bf 48}, 844 (1993)
[hep-th/9301047].
 }

\lr \tsnp{ 
A.~A.~Tseytlin,
``Effective action of gauged WZW model and exact
string solutions,''
Nucl.\ Phys.\ B {\bf 399}, 601 (1993)
[hep-th/9301015].
 ``Conformal sigma models corresponding to
gauged Wess-Zumino-Witten theories,''
Nucl.\ Phys.\ B {\bf 411}, 509 (1994)
[hep-th/9302083].
}

\lref\barsf { I. Bars and K. Sfetsos, \jnl \pl,  B277, 269,  1992;
 \jnl \pr,  D46,  4495, 1992.}

\def \t {\tau}

\lr\louk{ R.~C.~Myers and J.~Z.~Simon,
``Black Hole Thermodynamics In Lovelock Gravity,''
Phys.\ Rev.\ D {\bf 38}, 2434 (1988).
;
gr-qc/9610071}

\def \V {{\cal V}}

\lr \polch{ J.~Liu and J.~Polchinski,
``Renormalization Of The Mobius Volume,''
Phys.\ Lett.\ B {\bf 203}, 39 (1988).
}

\lref \york{J.~W.~York,
``Role Of Conformal Three Geometry In The Dynamics
Of Gravitation,''
Phys.\ Rev.\ Lett.\ {\bf 28}, 1082 (1972).
}

\lr\area{C.~G.~Callan, R.~C.~Myers and M.~J.~Perry,
``Black Holes In String Theory,''
Nucl.\ Phys.\ B {\bf 311}, 673 (1989).
R.~Myers,
``Superstring Gravity And Black Holes,''
Nucl.\ Phys.\ B {\bf 289}, 701 (1987).
}

\lr \sug {L.~Susskind and J.~Uglum,
``Black hole entropy in canonical quantum gravity
and superstring theory,''
Phys.\ Rev.\ D {\bf 50}, 2700 (1994)
[hep-th/9401070].
 }

\lr \mye{ R.~C.~Myers,
``Higher Derivative Gravity, Surface Terms And
String Theory,''
Phys.\ Rev.\ D {\bf 36}, 392 (1987).
}

\lr\wald {  M.~Visser, 
``Dirty black holes: Entropy as a surface term,''
Phys.\ Rev.\ D {\bf 48}, 5697 (1993)
[hep-th/9307194].
V.~Iyer and R.~M.~Wald,
``Some properties of Noether charge and a proposal
for dynamical black hole entropy,''
Phys.\ Rev.\ D {\bf 50}, 846 (1994)
[gr-qc/9403028].
}

\lr \dewi{  9910179  Cardoso, de wit, Mohaupt}

\lr\furs{D.~V.~Fursaev and S.~N.~Solodukhin,
``On the description of the Riemannian geometry in
the presence of conical defects,''
Phys.\ Rev.\ D {\bf 52}, 2133 (1995)
[hep-th/9501127].
}

\lr\tsone{ A.~A.~Tseytlin,
``On the form of the black hole solution in D = 2
theory,''
Phys.\ Lett.\ B {\bf 268}, 175 (1991).
}

\def \dM {{\del \M}}

\lr\tyut{
R.~E.~Kallosh, O.~V.~Tarasov and I.~V.~Tyutin,
``One Loop Finiteness Of Quantum Gravity Off Mass Shell,''
Nucl.\ Phys.\ B {\bf 137}, 145 (1978).
 }

\lr \hort{ G.~T.~Horowitz and A.~A.~Tseytlin,
``On exact solutions and singularities in string
theory,''
Phys.\ Rev.\ D {\bf 50}, 5204 (1994)
[hep-th/9406067].
}

\lr\bhtwoentr{ J.~D.~Creighton and R.~B.~Mann,
``Quasilocal thermodynamics of two-dimensional
black holes,''
Phys.\ Rev.\ D {\bf 54}, 7476 (1996).
R.~C.~Myers,
``Black hole entropy in two-dimensions,''
Phys.\ Rev.\ D {\bf 50}, 6412 (1994)
[hep-th/9405162].
S.~N.~Solodukhin,
``Two-dimensional quantum corrected eternal black
hole,''
Phys.\ Rev.\ D {\bf 53}, 824 (1996)
[hep-th/9506206].
}

\lr \frol { V.~P.~Frolov,
``Two-dimensional black hole physics,''
Phys.\ Rev.\ D {\bf 46}, 5383 (1992).
}

\lr \alex{ H.~Liebl, D.~V.~Vassilevich and
S.~Alexandrov,
``Hawking radiation and masses in generalized
dilaton theories,''
Class.\ Quant.\ Grav.\ {\bf 14}, 889 (1997)
[gr-qc/9605044].
}

\lr \tone{ A.~A.~Tseytlin,
``Renormalization Of Mobius Infinities And
Partition Function Representation For String Theory
Effective Action,''
Phys.\ Lett.\ B {\bf 202}, 81 (1988).
}

\lr\KAZ{ V.~A.~Kazakov,
``Solvable matrix models,''
hep-th/0003064.
}

\lr\BULKA{ D.~Boulatov and V.~A. Kazakov,
``One-dimensional string theory with vortices as the upside down
matrix oscillator,''
Int.\ J.\ Mod.\ Phys.\ A {\bf 8}, 809 (1993)
[hep-th/0012228].
}

\lr\SAL{ S.~Yu.~Alexandrov, private communication. }

\lr\DJP{S.~R.~Das and A.~Jevicki,
``String Field Theory And Physical Interpretation Of D = 1
Strings,''
Mod.\ Phys.\ Lett.\ A {\bf 5}, 1639 (1990).
J.~Polchinski, ``What is the string theory?'', hep-th/9411028.
}  

\lr \osbornetc { C.~G.~Callan, E.~J.~Martinec,
M.~J.~Perry and D.~Friedan, ``Strings In Background Fields,'' Nucl.\
Phys.\ B {\bf 262}, 593 (1985).
A.~A.~Tseytlin,
``Sigma Model Weyl Invariance Conditions And
String Equations Of Motion,''
Nucl.\ Phys.\ B {\bf 294}, 383 (1987).
R.~R.~Metsaev and A.~A.~Tseytlin,
``Order Alpha-Prime (Two Loop) Equivalence Of The
String Equations Of Motion And The Sigma Model Weyl
Invariance Conditions: Dependence On The
Dilaton And The Antisymmetric Tensor,''
Nucl.\ Phys.\ B {\bf 293}, 385 (1987).
C.~M.~Hull and
P.~K.~Townsend,
``String Effective Actions {}From Sigma Model
Conformal Anomalies,''
Nucl.\ Phys.\ B {\bf 301}, 197 (1988).
H.~Osborn,
``String Theory Effective Actions {}From Bosonic
Sigma Models,''
Nucl.\ Phys.\ B {\bf 308}, 629 (1988).
 }

\baselineskip8pt
\Title{\vbox
{\baselineskip 6pt
{\hbox {LPTENS-01/22}}{\hbox{OHSTPY-HEP-T-01-008 }}
}}
{\vbox{
\centerline {On  free energy  of    2-d black hole}
\medskip
 \centerline { in bosonic string theory  } 
\vskip4pt }}
\centerline  {V.A. Kazakov$^{1}$\footnote {$^*$} 
{e-mail address:  vladimir.kazakov@lpt.ens.fr  } 
and 
A.A. Tseytlin$^{2}$\footnote{$^{\star}$}
{\baselineskip8pt e-mail address: tseytlin.1@osu.edu}\footnote{$^{\dagger}$}{\baselineskip8pt
Also at  Imperial College,  London and  Lebedev  Physics
Institute, Moscow.} 
}
\medskip
\smallskip\smallskip
\centerline {$^1$\it  Laboratoire de Physique
Th\'eorique de l'Ecole Normale Sup\'erieure }
\smallskip
\centerline {\it 24 rue Lhomond, 75231 Paris Cedex,   France }
\medskip
\centerline {$^2$\it  Physics Department, Ohio State University}
\smallskip
\centerline {\it  Columbus, OH43210-1106, USA}
\medskip
\bigskip\bigskip
\centerline {\bf Abstract}
\bigskip
\baselineskip6pt
\noindent
Trying to interpret recent matrix model results (hep-th/0101011) we
discuss computation of classical free energy of exact dilatonic 2-d
black hole from the effective  action of string theory.  The euclidean
space-time action evaluated on the black hole background is divergent
due to linear dilaton vacuum contribution, and its finite part depends
on a subtraction procedure.  The thermodynamic approach based on
subtracting the vacuum contribution for  fixed values of temperature and
dilaton charge at the ``wall" gives (as in the leading-order black
hole case) $S= M/T $ for the entropy and  zero value for the free
energy $F$.  We suggest that in order to establish a correspondence
with a non-vanishing matrix model result for $F$ one may need an
alternative reparametrization-invariant subtraction procedure using
analogy with non-critical string theory (i.e. replacing the spatial
coordinate by the dilaton field). The subtraction of the dilaton
divergence then produces a finite value for the free energy.  We also
propose a microscopic estimate for the entropy and energy of the black
hole based on the contribution of non-singlet states of the matrix
model.

 \medskip 

\Date{April 2001}

\def \lo {{leading-order\ }}

\def \bh {{black hole\ }}
\noblackbox \baselineskip 16pt plus 2pt minus 2pt


\newsec{Introduction}
Recent investigation \kkk\ suggested a way to compute
free energy of the dilatonic 2-d \bh \refs{\witt,\rabi}
using matrix model methods.
It is of obvious interest   to try to compare 
the matrix model  prediction with  the free energy which follows 
from the string low-energy space-time effective action.

The previous analysis of thermodynamics of euclidean black hole \gp\
used the leading-order (1-loop) background of \rabi\ with the
conclusion that the entropy is $S= M/T$, implying that the black-hole
free energy $F = E- S T$ is zero (equivalent results were found in
\refs{\nap,\frol,\bhtwoentr}).  Here $E=M$ is the \lo \bh mass (equal 
to its ADM mass \refs{\witt,\gp,\nap}), i.e. the full energy minus the
divergent dilaton vacuum contribution.

At the same time, the bosonic matrix model calculation \kkk\ gave a
non-zero result 
\eqn\FINF{   -{\cal F}_0 = \beta F \propto   e^{-2\phi_0}  }
for the tree-level part of the string theory partition function.
One may try to reconcile these facts by acknowledging that since the
full string-theory expression contains the divergent dilaton vacuum
contribution, a finite (subtracted) value of free energy may be
ambiguous, i.e. non-universal \kkk.  The finite matrix model result
was obtained by solving Toda equation with certain natural boundary
conditions.  On string theory (effective gravitational field theory)
side, one finds a finite (zero) value for $F$ after a particular
(thermodynamic ensemble motivated) 
subtraction of the dilaton vacuum contribution.
 
The comparison between string theory and matrix model results in \kkk\
was referring to the analysis of 2-d \bh thermodynamics for the
leading-order form of the black hole background.  One may expect that
the effective field theory result $F=0$ may change if one considers
the exact (all-order in $\a'$) form \dvv\ of the \bh background.
For example, one may finish with $S = \nu M/T$, where $\nu$ is a
numerical coefficient different from one, leading to $F\not=0$.
 
The aim of the present paper is to address this issue.  First, we
shall repeat the thermodynamic analysis of \gp\ replacing the
leading-order black hole background by its exact form.  We shall show
that  the subtracted black-hole part of the free energy is still zero.
Our discussion will highlight several important open questions, in
particular, the origin of boundary terms in string theory effective
action and the meaning and consistent implementation of subtraction of
the dilaton vacuum contribution.

The approach of \gp\ to computing the free energy is based on the
assumption that one can prepare the canonical ensemble (for the black
hole in equilibrium with radiation) with any values of the temperature
and dilaton charge at the boundary. At the same time, the value of the
temperature of the 2-d dilatonic  black hole background in string theory
is not a free parameter being fixed by the central charge condition.
The thermodynamic approach is thus implicitly assuming that this
constraint can be relaxed, i.e. that the temperature can be made
arbitrary by adding extra matter fields.

This approach may not, however, be an adequate one for comparison with
the matrix model.  We shall suggest that in this case one should use a
different subtraction procedure in defining the effective action.  As
a result, we shall find a non-zero value for the string partition
function on 2-sphere consistent with the matrix model prediction \kkk.
  

While in the case of the fermionic string (with $N=1$ world-sheet
supersymmetry) the leading-order 2-d \bh solution of
\refs{\rabi,\witt}\ is exact, it receives, in general,
$\a'$-corrections in the bosonic case.  Below (in section  3)  we shall
reconsider the computation of the classical contribution to the \bh
free energy using the $\a'$-corrected form of the background \dvv\
(see also \rf{ \bsfet,\tsnp}). This background is the exact solution
of the conformal invariance equations to all orders in sigma model
loop expansion in the standard (dimensional regularization with
minimal subtraction) scheme \refs{\tsone,\panv,\tspl}.\foot{There
exists a formal scheme in which the leading-order solution remains
exact \refs{\tspl,\ttt}, but it is the exact background which is
``seen" by a string (tachyon) probe \dvv.}  We shall follow closely
the discussion in \gp\ (see also
\refs{\nap,\frol}).  Their results showed analogy with similar results
in four dimensions \GH\ but the presence of the dilaton field
introduced important subtleties.

The question of computing the entropy for the exact black hole
background of \dvv\ was raised already in \gp, but it was not clear
how to do that computation given that the free energy should be
determined by the value of the full  Euclidean string action $I$ 
(with  unknown higher order
$\a'^n$ corrections)
 evaluated on
the exact solution.\foot{For discussions of \bh entropy in higher
derivative gravity theories see, e.g.,
\refs{\area,\wald,\rrbh,\furs}.  } 

In spite of the fact that the higher order $\a'$ terms in $I$ are not
known in general, one of our main observations is that the value of
the full action {\it at the exact conformal point}, i.e. its exact
extremum, may be computed, provided one makes a natural assumption
about the structure of the boundary terms in the action.  As we shall
argue in section 2, the (properly defined) volume term in the
effective action always vanishes at its exact extremum, and there
exists a natural simple choice for the boundary term.

In section 3 we shall first review the structure of the exact 2-d
\bh background \dvv\ and then  compute the value of the  boundary term 
in the action on this background. 
In section 3.1 we shall find the corresponding free energy $F$ and
entropy $S$ by introducing an IR cutoff (``position of a wall") and
treating, following \gp, the dilaton charge and the temperature at the
wall as the arguments of $F$.  Subtracting the dilaton vacuum
contribution from the energy and thus getting the \bh mass $M$ we
shall then find that $S= M/T$, implying that the free energy vanishes.
We shall then explain in section 3.2 an alternative (non-critical string theory motivated)
subtraction procedure for the string effective action on the black
hole background. It   leads to a finite value
of the action  which is qualitatively
consistent with the expression in \kkk.  Alternative ways of doing the
infinite subtraction of the dilaton vacuum contribution 
 will be discussed  in section 3.3.

In section 4 we shall review the matrix model result of \kkk\ for the
free energy in the tree approximation and try to extract from it the
energy and the entropy of the black hole
(related matrix model Hamiltonian  will be given in Apendix). 

Some concluding remarks will be made in  section 5.

\newsec{Bulk and  boundary terms in  
string theory  effective action}

The standard way to compute a classical entropy associated with a
gravitational background is to evaluate the Euclidean effective action
on this solution \GH.  One follows the analogy with path integral
approach to quantum gravity where the partition function is given by
the integral over the space of all field configurations that are
periodic in Euclidean time. The classical contribution is thus $
e^{-I} = e^{-\beta F} $.

The string effective action for ``massless" modes should be given, in
general, by the sum of a bulk term $I_\M$ and a boundary term $I_{\del
\M}$.  Evaluating it on a solution following \GH\ one gets the free
energy
\eqn\fef{  F= T \  I   \ , \ \ \ \ \  \ \ \   I= I_\M + I_{\del \M} \ , 
\ \ \ \ \  \ \ \   T = \b^{-1}  \ . }
Both terms in $I$ are expected to be infinite series in $\a'$ and thus
are not known exactly.

We shall by-pass this problem by arguing that a properly defined bulk
term $I_\M $ should actually vanish when evaluated on the exact
conformal background, while the corresponding boundary contribution
$I_\dM $ in this case should be given just by the two standard
lowest-derivative metric and dilaton dependent terms.  If the explicit
all-order form of a conformal background is known, one is then able to
get the exact expression for the free energy $F$ just by evaluating
the two leading boundary terms in $I$ on this background.

Let us consider first the bulk term $I_\M$.  To obtain the effective
action for the massless closed string metric and dilaton fields
$I[G,\pp]$ from the string partition function $Z$ (generating
functional for correlators of vertex operators \FT) one needs to
subtract both the massless poles (UV logarithms) \TT\ and the
$SL(2,C)$ M\"obius infinities \refs{\mec}.  While the former are
power-like in the open string case \refs{\tone,\polch}, they are also
logarithmic in the closed string case \refs{\polch,\mec}, implying
that an ``extra" log should be subtracted from $Z$.  The RG invariant
way to do that subtraction is to apply $\del \ov \del \ln\ep$ to the
bare value of $Z$ \mec.  Expressed in terms of renormalized couplings
(tuning the renormalized value of the tachyon coupling to zero) the
effective action is then given by
\rf{\mec,\mecc}\  ($\l^i = (G, \pp)$) 
 \eqn\eeqq{ I_\M = - ({\del Z \ov \del \ln \ep})_{{\ep=1}} =
 \beta^i\cdot { \delta Z \ov \delta \l^i} \ , } 
where the objects in the second equality are assumed to be expressed
in terms of the renormalized couplings.  One can then argue that there
exists a scheme where \rf{\mec,\mecc}
\eqn\voot{ I_\M= - \big(
 { \del V \ov \del \ln \ep } \big)_{{\ep=1}} \ , \ \ \ \ \ \ \ \ \ \ \
 V = \k_0 \int_\M d^D x\ \V \ ,\ \ \ \ \ \ \V\equiv
\sqrt G  \  e^{-2\pp  }  \  . }
Here $G$ and $\p$ in $\V$ are the running (cut-off dependent)
couplings.  Using the diffeomorphism invariance of $Z$, the expression
for $I_\M$ can be rewritten also as $ I_\M = \bar \beta^i\cdot {
\delta Z \ov \delta \l^i}$, where $\bar \b^i = \b^i + (\delta
\l^i)_{\a' \del \pp} $ 
are the Weyl-anomaly coefficients, 
which differ from the $\b$-functions by the diffeomorphism terms, 
i.e.  $\bar \b^\pp = \b^\pp + \a' \del^m
\pp\del_m \pp, \ \ \bar \b^G_{mn} = \b^G_{mn} + 2\a' \cD_m\cD_n \pp $.
Explicitly ($\k_0 \sim \a'^{-D/2}$)
\eqn\nuew{
 I_\M = \k_0 \int_\M d^D x\ \sqrt G\ e^{-2\pp }\ \big ( 2 \b^\pp - \ha
 G^{mn} \b^G_{mn} \big) = 2 \k_0 \int_\M d^D x\ \sqrt G\ e^{-2\pp }\
 \td \b^\pp \ , }
where the ``central charge Lagrangian"
\eqn\coo{  \td \b^\pp  = 
\bar \b^\pp -  { 1 \ov 4}  G^{mn} \bar \b^G_{mn} \  } 
is the total conformal anomaly of the 2-d \sm (coinciding with the
central charge at the conformal point): $<T^a_a> = { 1 \ov 4\pi} \td
\b^\pp R^{(2)} + O(\bar \beta^G) $.  The leading terms in its
derivative expansion are
\eqn\bbb{
 \td \b^\pp = 
c_0 -  { 1 \ov 4} \a'( R  +  4 D^2 \pp -  4 \del_m \pp \del^m \pp)
 + O(\a'^2 R^2_{mnkl})\ , 
\ \ \ \ \ \   c_0  =   {{ 1 \ov 6} }(D- 26)\ . }
The resulting action reproduces the Weyl-invariance conditions $\bar
\beta^i=0$ as its equations of motion \rf{\osbornetc}.

Even though we do not know the explicit form of the ``central charge"
$ \td \b^\pp $ to all orders in $\a'$, we can still deduce that since
$ \td \b^\pp=0$ is one of the conditions of 2-d Weyl invariance
(dilaton equation of motion), the bulk term in the action given by
\nuew,\bbb\ {\it vanishes} on any exact string solution.

This is the expected conclusion: the partition function of a 2-d CFT
should not depend on a cutoff, and thus its derivative in \eeqq\ (or,
equivalently, its ratio with the infinite volume of the $SL(2,C)$
M\"obius group one needs to divide over to obtain the {\it string
theory} partition function on $S^2$) should {\it vanish} at a
conformal point.

Let us now turn to the boundary contribution $I_{\del \M }$. While we
do not know how to derive it systematically from the string \sm path
integral, one can try to fix it using indirect considerations.  The
two leading field-dependent terms in \bbb\ -- $R$ and $ D^2 \phi$ --
contain second derivatives, so we may follow the standard logic of
``integrating by parts", i.e.  adding the corresponding boundary terms
to have a well-posed variational principle that involves only the
fields and their first derivatives \rf{\york,\GH}.  That gives
\eqn\boun{
I_{\del \M } = - 2 \a' \k_0 \int_{\partial {\cal M}} d^{D-1} x \sqrt
  \g \ e^{ -2 \pp} (K - 2 \del_n \phi) = - 2 \a' \k_0 \int_{\partial
  {\cal M}} d^{D-1} x \sqrt \g \ \nabla_a ( e^{ -2 \pp} n^a) \ , }
where $ \del_n = n^a\nabla_a $ ($n^a$ is the unit outward normal
vector to the boundary), $K$ is the second fundamental form of the
boundary $\nabla_a n^a$ and $\g$ is the induced metric.  For a
diagonal metric of the type we shall consider below this can be put in
the following simple form
\eqn\bon{
I_\dM  = - 2 \a' \k_0 \int_{\dM}  d^{D-1} x \  \del_n
 ( \sqrt G  e^{ -2 \pp} ) 
=  - 2 \a'\k_0  \int_{\del \M}   d^{D-1} x\   \del_n \V \ , } 
where $\V$ is the same volume density as in \voot.

Naively, one could think that the presence of higher-order terms in
the beta functions and the effective action implies that one should
also add higher-derivative terms to the boundary part of the action.
However, in general it is not possible to derive such terms just from
the requirement of having a consistent initial value problem
\mye.\foot{In the standard QFT context higher-derivative boundary
terms are indeed induced in the quantum effective action if one
quantizes matter fields coupled to gravity on a manifold with
boundary.  Heuristically, one could try, by analogy with how that is
done for the Seeley coefficients of Laplacians, to derive the boundary
terms which are associated with higher-order ($R^n+...$) terms in
$I_\M$ by replacing $\M$ by its double with no boundary, and then
computing $I_\M$ for the fields on the double that will have
delta-function-type singularities along the joining (boundary)
submanifold, producing boundary terms as a result.  However, there
does not seem to be an intrinsic reason for this procedure in the
present context.  Again, the idea of reconstructing boundary terms
from a well-definiteness of variational problem does not seem to apply
to higher-order terms (one would in any case need to fix all higher
derivatives of variation of the fields at the boundary).}

Our main assumption will be that (at least at the conformal point)
\boun\ represents the {\it exact} expression for the boundary term in
the tree-level string effective action.  This is a natural assumption
for the following reason.  If one defines the string effective action
by reconstructing it from the string scattering amplitudes on an
asymptotically flat space-time, one needs to properly define the 2-nd
derivative (propagator) terms in the action -- by adding the boundary
terms \boun.\foot{Note that the classical graviton S-matrix in the
Einstein theory is indeed generated by the boundary term $K$ in \boun\
evaluated on asymptotic wave $h^{(in)}$ dependent solution of the
Einstein equations \tyut.}  One may then think that all higher-order
corrections (that can be always arranged so that they do not change
the propagators of the massless fields
\refs{\tsred,\grwi})  
should represent  interaction vertices (treated  perturbatively
in $\a'$) and   thus should not 
be accompanied by   boundary terms.

\newsec{Free energy of exact 2-d black hole  background}

The 2-d \bh background \rabi\ was shown in \witt\ to be described by
exact CFT represented by gauged $SL(2,R)/U(1)$ WZW model.  The exact
background as ``seen" by a local fundamental string probe (tachyon
field) was extracted by
\dvv\  from  the Hamiltonian   of this  CFT
interpreted as a generalized Laplacian on the coset space.\foot{The
same background can be obtained also directly in the \sm framework by
integrating out 2-d gauge fields \refs{\bsfet,\tsnp} (see \ttt\ for a
review). } It was checked that this background is indeed a solution of
the $\bar \beta^i=0$ equations at the 2-loop \tspl\ and 3-loop \panv\
levels (with the beta-function computed in the standard dimensional
regularization with minimal subtraction scheme).

The expression for a background extracted from a gauged WZW model is,
in general, scheme-dependent \rf{\tspl,\ttt}.  There exist a
non-standard ``leading-order scheme" in which the background has its
leading-order 1-loop form (but the tachyon beta-function is modified
by $\a'$ corrections) and a ``CFT scheme" where (in bosonic case) the
background gets corrections to all orders in $\a'$ (but the tachyon
equation retains its leading-order form).\foot{Note that in 2
dimensions one can always choose a scheme such that any solution of
the effective equations has its leading-order form \hort.}  The two
backgrounds describe the same string geometry in the sense that since
a local probe of the geometry is the tachyon field, and the tachyon
equation remains exactly the same differential equation as implied by
the coset CFT (even though it looks different when expressed in terms
of different couplings $G$ and $\p$).

 The  Euclidean 2-d \bh background in the leading-order scheme  
 \refs{\rabi,\witt} 
\eqn\exn{ ds^2_{lead} =  
 dx^2 + \ { {\tanh}^2 bx } \ d{\t}^2 \ , \ \ \ \ \ \ \p_{lead} = \p_0
- {\ln \ch } bx \ , }
is related to the ``exact" background in the CFT scheme \dvv\foot{In
our normalization the \sm action is: $I= {1\over { 4 \pi \a' }} \int
d^2 z \sqrt {g} \ [ \del_a x^m
\del^a x^n G_{m n}(x)\   +   \a' R^{(2)}  \p (x) ] $.
 The coordinates that naturally appear in the WZW context are: $ r= b
x, \ \vp= b \t$, \ $\a'b^2= { 1 \ov k-2}$. }

\eqn\exann{ ds^2 = dx^2 + \ { {\tanh}^2 bx \ov 1 \ - {\ p \ }
{\tanh}^2 bx }   \ d{\t}^2   \ ,  \ \ \  }  \eqn\eeee{
 \p = \p_0  -  {\ln \ch }  bx   -
\four  \ln ( 1 \ - \ p \ {\tanh}^2 bx )\ ,  }
\eqn\exxxx{  \  
\a'b^2 = {1 \ov k-2}\ ,\  \ \ \  \ \ \ \ 
p\equiv { 2 \a' b^2 \ov 1 + 2 \a'b^2 } = {2\ov k} \ , \ \ \ \ \ \
\ \ \ \ \ 
D - 26  + 6\a'b^2  =0  \ , }
by the following local covariant redefinition \tspl
\eqn\cftf{   ({G_{mn}})_{lead}  = G_{mn}  - { 2\a'  \ov 1  + \ha \a' R  }  [
\del_m \p \del_n  \p  -    (\del \p)^2  G_{mn}]
\ , }
$$ \p_{lead}=\p - \four \ln ( 1 + {\textstyle {1\ov 2}} \a' R ) \ .
$$ Note also that\foot{Such equality of the measure factors is true in
general for backgrounds obtained from gauged WZW models
\refs{\bsfet,\tsnp}.}
\eqn\mer{
\sqrt {G_{lead}}\ e^{-2 \p_{lead}} = \sqrt G\ e^{-2 \p}  \ . }
 $e^{\p_0}= g_s$ is the string coupling at the horizon $x=0$
(tip of the ``cigar"). 

To find the free energy we shall use the exact form of the background
\exann.  Since $\td \beta^\p$ (computed in the standard scheme)
vanishes for the exact background \exann,\eeee, the bulk term in the
action \nuew\ is zero, and thus the free energy \fef\ is determined by
the boundary term \boun. In view of the representation
\bon\ and  the equality of the two volume density 
factors \mer, the boundary contribution \boun\ evaluated for the exact
background will look formally the same as for the leading solution
computed in \gp.  However, the thermodynamic properties of the two
backgrounds differ since they have different large distance
asymptotics.

 The  metrics \exn\ and \exann\ are of the  form
\eqn\class{
ds^2 = dx^2 + f^2(x) d \t^2 \ , }
for which  \boun\ reduces to \bon\   and  thus gives   
(in what follows we set 
$\a'\k_0= 1$)
\eqn\boni{
I_\dM  = - 2  \int_0^{\b_0}  d \t  \  \del_x
 ( f  e^{ -2 \pp}   ) |_{x= x_{bndry}} \ . } 
As a result, we  get  the  same-looking   expression 
for the action   as  found 
in \gp\  for \exn)  
\eqn\bonk{
I_\dM  = - 2 \b_0\  b \  e^{-2 \p_0}\  \ch 2 b x_{bndry}
  \ . } 
Here $\b_0$ is the period of $\t$ which is determined  from the condition of the absence of
 conical singularity near $x=0$.\foot{As usual, 
the flat space   has  topology of a cylinder,
$S^1 \times R^1$, 
where the Euclidean time is
periodic with an arbitrary period,  while the Euclidean black hole
has topology of a hemisphere, so  its  
local temperature is given by the inverse proper periodicity
of the Euclidean time needed to obtain a regular  
disc metric near the tip.}
 {}From \eeee\ we get (taking $x\to 0$) 
\eqn\temp{
T_0 = \b_0^{-1}  =  { b \ov  2\pi } =
  {  1 \ov 2 \pi \sqrt { \a'}  \sqrt{ k-2}   } \ .  }
  As a consequence, \bonk\ becomes
  \eqn\bank{
I_\dM  = - 4 \pi \  e^{-2 \p_0}\  \ch 2 b x_{bndry}
  \ . }
Note that while for the leading-order background
 \exann\
$f(x) \to 1$  for large $x$ and thus 
the asymptotic temperature is the same as 
 $T_0$, this is not true for the exact  background
 \exann: here $f(x) \to  (1-p)^{-1/2}$  and thus 
 the Hawking temperature  is   \pert\foot{One could, of course,
 avoid introducing $T_0$ 
by  rescaling $\tau$ in the metric \exann\ so that 
to make \class\
 having the  canonical flat space form  at infinity with $f\to 1$.
 Then $T$  would be the inverse period of the Euclidean
 time.}
 $$ 
  T\ =  \  \b^{-1}= 
{b \ov  2\pi}  (1- p)^{1/2}  = T_0 (1- p)^{1/2} $$
\eqn\asy{
= 
 {T_0 \ov  \sqrt{ 1 + 8 \pi^2 \a' T^2_0}}= 
  { 1 \ov  2\pi  \sqrt { \a'}  \sqrt k  }
  = { \sqrt p \ov 2\pi  \sqrt 2 \sqrt { \a'} } \ . \  }
Below we shall keep the dependence on the parameter 
$p$  in \exann,\eeee\ explicit so that the quantities corresponding to the leading-order black hole can be recovered 
  by setting $p=0$ (see \exxxx).
Note that if we put  $D=2$  in \exxxx\ 
we get $\a' b^2 =4$, i.e.  $k={9\ov 4}$ \witt\   and thus 
the temperature will not be a free parameter. 
It is possible to   assume  instead that the 
 temperature  may  be changed  by  
changing the effective central charge  as a result of 
 varying the  number   $N (=D-2$)
of extra massless ``matter"  fields  that can be added to the system.


\subsec{\bf  Thermodynamic ``subtraction at  the wall" approach}

As usual, we shall restrict $x$ to the interval $0\leq x < \infty$.
Following \gp, let us compute all thermodynamic quantities not
directly at the boundary $x=\infty$ but at a finite distance $x=x_\w$
(``position of a wall") treating this as an IR regularization. We
shall then subtract the divergent contribution of the asymptotic
cylinder.
This regularization is needed 
since   one does not have a translationally 
invariant   vacuum at infinity: 
there is a linearly growing dilaton.

The value of the temperature at the wall is determined by the
periodicity of the effective Euclidean time coordinate there, i.e.
\gp\ ($\t$ in \exann\ has the period $\b_0$, see \temp) 
\eqn\tmn{ T_\w
= T(x_\w)= T_0 f^{-1} (x_\w) = T_0 ( \coth^2 b x_\w - p)^{1/2} \ , }
or, equivalently, \eqn\tmno{ T_\w = T \bigg({ \coth^2 b x_\w - p \ov
1- p} \bigg)^{1/2}
\ ,   \ \ \ \  \ \ \ \ \ \ 
 T = T_\w (x_\w=\infty) \ .  } 
Note that as in the case of the
leading-order background with $p=0$ one has $T_\w \geq T$.

As in \gp\ we shall view the values of the temperature $T_\w= T(x_\w)$
and the dilaton $\p_\w=\p(x_\w)$ as the basic free parameters of the
thermodynamic system.

We define the dilaton charge as in \gp   
\eqn\dil{
\D_\w = e^{-2 \p_\w} = e^{-2 \p_0}  \cosh^2 bx_\w \ 
 (  1 \ - {\ p \ }{\tanh}^2 b x_\w )^{1/2} \ . } 
Then the free energy at the wall 
 that follows from \bonk\  is 
$$F_\w  =  T_\w I_\dM (x_\w) =  - 4\pi T_\w   e^{-2 \p_0} \cosh 2 b x_\w  $$
\eqn\fre{
=  - 4 \pi \D_\w \ ( \sqrt{ T^2_\w + p T^2_0}
 + { T^2_0 \ov  \sqrt{ T^2_\w + p T^2_0 }}  )  \equiv  F(\D_\w, T_\w) \ . }
For $p=0$ this reduces to the expression for the leading-order 
background \exn\ in \gp.
Note that  the free energy is divergent 
for $x_\w\to \infty$
\eqn\exrt{
F_\w =  - 2 \pi T e^{-2 \p_0}[ e^{  2 b x_\w}
  + 2  (1-p)^{-1}  + O( e^{-b x_\w}) ]  \ . }
One may try to subtract the divergence in a naive way, getting a {\it
non-zero} result for the free energy (but zero value of the energy,
see subsection 3.3 below).  However, such a direct subtraction of the
$e^{ 2 b x_\w}$ term is problematic since it is coordinate-dependent.

One systematic approach is to follow \refs{\gp,\frol}\ and to subtract
the flat space vacuum contribution to the free energy at fixed values
of the basic variables $T_\w$ and $\D_\w$.  The flat space metric is
\class\ with $f=1$ and $\p= \p'_0 - b x$, so that
\boni\ gives\foot{We take the flat space metric as $dx^2 +
d\tau'^2$ where $\tau'$ has periodicity $\beta_\w$.  An equivalent
 approach is to take it as the asymptotic form of \exann, $dx^2 +
 (1-p)^{-1} d\tau^2$ with periodicity of $(1-p)^{-1/2} \tau$ still
 being $\beta_\w$ to match the definition of temperature in \tmn.
 Note also that, as in \gp, we ignore the contribution of the other
 boundary of the flat-space vacuum cylinder.}  $ I_\dM^{(vac)}= -4
 \b_\w b \D_\w$, where we took the period of $\tau$ to be $\b_\w$ and
 replaced $e^{-2\p}$ by $\D_\w$.  Then
\eqn\cav{
F^{(vac)}_\w =  T_\w I_\dM^{(vac)}=  - 4   b  \D_\w = - 8 \pi T_0  \D_\w  \ , }
so that  the subtracted  value of the free energy is 
\eqn\suub{
F'_\w =  F_\w  - F^{(vac)}_\w  =
  - 4 \pi \D_\w \ ( \sqrt{ T^2_\w + p T^2_0}
 + { T^2_0 \ov  \sqrt{ T^2_\w + p T^2_0 }}    -  2 T_0 )  \ . } 
Expanding this at $x_\w \to \infty$ 
we find
\eqn\subb{
F'_\w =  - 4 \pi T  e^{-2 \p_0}  e^{-2 b x_\w} + 
O(  e^{-4 b x_\w}) \ , \ \ \ \ {\rm i.e. } \ \ \ \
 F'_\infty  =  0 \ .  }
Thus the subtracted value of the  free energy 
{\it vanishes}.


The  basic thermodynamic relations are
\eqn\rell{
S_\w = -\biggl({\partial F_\w\over \partial T_\w}\biggr)_{\D_\w} \ , \
\ \ \ \ \
\psi_\w = -\biggl({\partial F_\w\over \partial \D_\w}\biggr)_{T_\w} \ , \ \ \ 
\ \ \ \  }
 \eqn\rella{ E_\w= F_\w + S_\w T_\w \ , } where $S$ is the entropy,
 $\psi$ is the chemical potential associated with the dilaton charge
 and $E$ is the total energy.  {}From \cav\
\eqn\cave{
S^{(vac)}_\w = 0 \ , \ \ \ \ \ \ \ \ \ E^{(vac)}_\w = F^{(vac)}_\w = -
4 b \D_\w \ . }
 Differentiating \fre\ over $T_\w $ we find
\eqn\sss{
S_\w =  4\pi \D_\w ( 1 + p  { T^2_0 \ov T_\w ^2 } )^{-3/2} 
[ 1 - (1-p) { T^2_0 \ov T_\w ^2 } ] 
=  4\pi \D_\w ( 1 + { p\ov 1-p}   { T^2 \ov T_\w ^2 } )^{-3/2} 
 ( 1 - { T^2 \ov T_\w ^2 } )   \ .  }
While the entropy was  constant for  the leading $p=0$ 
solution \gp,  this is no longer so for the exact solution. 
Explicitly, in terms of $x_\w$ 
\eqn\sss{
S_\w =  4\pi e^{-2 \p_0}  (  1 \ - {\ p \ }{\tanh}^2 b x_\w ) \ , }
so that  the entropy   is positive as 
long as  $ p < 1$,  i.e. $ k-2 > 0$.
Its  asymptotic $x_\w \to \infty$ 
 boundary value is\foot{It is non-zero 
  since   the vanishing of  the factor 
$1 - { T^2 \ov T_\w ^2 }$ in \sss\ at $x_\w=\infty$ 
 is compensated by the divergence of the dilaton factor.}
\eqn\ses{
S  =  4\pi e^{-2 \p_0}  (1-p)  \ . } 
For the energy \rella\ 
we find
$$
E_\w = -  4 \pi \D_\w  { T^2_0 \ov T_\w} ( 1 + p  { T^2_0 \ov T_\w ^2 } )^{-3/2} 
[ 2  + p +   p (p+1) { T^2_0 \ov T_\w ^2 }  ] 
$$
\eqn\ene{
= \ -  4\pi e^{-2 \p_0}  T_0  ( \sinh 2 b x_\w + p  \tanh bx_\w  )  
 (  1 \ - {\ p \ }{\tanh}^2 b x_\w )^{1/2} \ .  }  
Like  the full 
free energy \exrt, this expression is divergent at large $x_\w$  
--  it should represent the combined energy of the background dilaton
field and the black hole.
Subtracting   the  vacuum 
 value  $E^{(vac)}_\w$ in \cave\  following \gp\  
we get 
$$E'_\w =  E_\w -  E^{(vac)}_\w 
=  8 \pi \D_\w  T_0 \bigg[1 -   { T_0 \ov T_\w} ( 1 + p  { T^2_0 \ov T_\w ^2 } )^{-3/2} 
\big( 1  + { 1\ov 2}  p +   (p^2+p) { T^2_0 \ov 2  T_\w ^2 }  \big) 
  \bigg] \ $$
\eqn\esu{
 =  4 \pi T e^{-2 \p_0}  (1-p)  +  O( e^{-2b x_\w}) 
\ . } 
The asymptotic value of the subtracted energy should 
represent   the  mass of the
 black hole,\foot{This expression 
 agrees, for $p=0$,  with the ADM mass of the leading-order black hole
 \refs{\witt,\gp,\nap}. One may expect that, like the free energy, 
  the 
  ADM mass of the exact black hole   should 
  not depend on the unknown details of the bulk term in the action,
  and should be given just by a boundary term.
  However, the expressions  for the ADM mass in 
 \refs{\witt,\gp,\nap} (see also \alex) 
 were derived using leading-order  form 
 of the action or equations of motion and  do
 not seem to apply to the present  exact background.
 In particular, the above result for $M$ 
 is different from what  follows from 
  the expression  $ M_{\rm ADM}= 2(e^{-2 \p} \del_x f )_{x=\infty}$
  in \gp\ for 
   the  background \exann,\eeee\
   (we correct a misprint in \gp: $f^2 \to f$;
   equivalent result was given  in \nap).
     It also disagrees with the 
   expression for the mass given  in  \pert.
   Using the definitions of the ADM mass in \refs{\gp,\nap}
   one finds 
   $M= 4 \pi T e^{-2 \p_0}  (1-p)^{-3/2} \ . $
   The definition in \refs{\witt}, i.e.  
   $M= 4 e^{-2 \p_\infty} \del_x \varphi$, 
   $\phi= \phi_\infty + \vp + O(e^{-4bx})$ (see also \alex) 
     gives
   $M= 4 \pi T e^{-2 \p_0}  (1-p)^{-1}  \ . $
   All these expressions agree for $p=0$ only. 
   }
 \eqn\suubb{ 
M= E' =  4 \pi T e^{-2 \p_0}  (1-p)   \ . } 
Expressing  the entropy \ses\ in terms of the mass we thus get
\eqn\ernt{
S = \beta  E'  =  { M \ov T}  
\ . }
This  is  the same relation  that was found in \gp\ 
for the leading-order \bh (and  is the 2-d analog of the 
4-d Schwarzschild black hole  relation 
$S= 4 \pi M^2$). 
It  is, of course, consistent with the vanishing 
of the subtracted value of the free energy in \subb.

\subsec{\bf  ``Dilaton  subtraction"  approach }

Given that the matrix model result \kkk\ for the tree-level free
energy is finite and {\it non-zero}, it is natural to try to look for
alternative prescriptions for computing $F$ that may also give a
non-vanishing result on the effective field theory side.  Let us go
back to the expression for the Euclidean action \bank\ and suggest a
subtraction procedure that seems naturally adapted to comparison with
the matrix model calculation.

The matrix model computation uses 
Sine-Liouville model (see section 4 below) 
so the  spatial direction   $x$  in \exann\     
    may be interpreted as a
``Liouville direction" $\vp$.
  Furthermore, it can be always traded for the
dilaton field $\phi$ which is the true counterpart of 
the Liouville field
$\vp$.\foot{For example, one can change coordinates in \exann,\eeee\ so
that the dilaton becomes a linear function of the new coordinate
$x'$.}  Let us then re-express the effective action \bank\ in terms of
the dilaton variable $\phi$ using
\eeee\
\eqn\IDIL{  I_{\partial\M}= -{4\pi\over 1-p}  e^{-2\p_0} 
\left[-1+\sqrt{p^2+4(1-p)e^{-4(\p-\p_0)}}\right]   .}
%
This expression contains a divergence corresponding to the infrared
behaviour $\phi\to - \infty$ in the target space.  This divergence may
be interpreted as an UV one on the world sheet of the Sine-Liouville 
model 
regarded as a non-critical string theory (if $\vp$ is associated
with the conformal factor of the 2-d metric then the covariant cutoff
corresponds to $ e^{-2 \vp} |\Delta z|^2 > 1/\Lambda^2 \to 0$).
Introducing a cutoff $\Lambda$ on $e^{-\phi } \to \infty $ (similar to
the cutoff used in the matrix model) we get from \IDIL\
\eqn\IDILE{  I_{\partial\M} \simeq 
-{8\pi\over (1-p)^{1/2}} \Lambda^2+{4\pi\over 1-p}
e^{-2\phi_0}- {\pi p\over (1-p)^{3/2}}e^{2\phi_0}{1\over
\Lambda^2}+O({1\over \Lambda^4})\  .}
As in the matrix model approach 
 discussed in section 4, it is natural to drop the
non-universal cutoff-dependent terms in \IDILE. We then  obtain the
following finite value for the string partition function\foot{ The
same expression for the finite value of the action \bonk\ was obtained
by S. Alexandrov using a subtraction of the flat space contribution
``at equal dilatons'' \SAL .
}
\eqn\IFIN{-\CF_0\equiv  I'_{ \partial\M  }= {4\pi\over 1-p} e^{-2\phi_0} \ .  }
 It has the same scaling $\CF\sim e^{-2\phi_0}\sim g_s^{-2}$ as the
matrix model result.  The coefficients are difficult to compare
systematically: for that one needs to determine several more orders in
the $g_s^2$ expansion.


\subsec{\bf Alternative subtractions}
One may look also for other 
alternative subtractions that may  reconcile 
the results of the matrix model and effective
field theory approaches.
One could expect that a subtracted field-theory expression for $F$
could be zero for the leading-order $p=0$ background but non-zero for
$p\not=0$, i.e.  for the exact background.  However, such a
prescription which is, at the same time, consistent with thermodynamic
relations does not seem to exist.
Let us  make few  remarks about other 
 subtraction recipes which
either lead to the zero value of the energy inconsistent with
thermodynamics or still give $F=0$ for any value of $p$.

First, let us go back to the expansion \exrt\ and subtract the
divergence in the naive (non-reparametrization invariant) way, thus
getting a {\it non-zero} result for the free energy
\eqn\naii{
F' =  - 4 \pi T e^{-2 \p_0}  (1-p)^{-1}
\ . } 
If one formally  computes the entropy 
according to $S_\infty = -
{\partial F_\w\over \partial T} $
one  then gets the result 
\eqn\sees{  S= 4 \pi e^{-2 \p_0} (1-p)^{-1} \ , }
which is different from \ses\ but is still in agreement with the
leading-order expression for the entropy once we set $p=0$:\ $S= 4
\pi (e^{-2 \p})_{horizon}= 4 \pi e^{-2 \p_0}$ \refs{\gp,\frol}.  The
problem, however, is that then $E' = F'+ S T $ (see \rella) is zero
instead of being equal to the black hole mass.

This is a general conclusion: if one considers the subtracted value of
the free energy to be a  finite linear function of the temperature
directly at infinity and varies $T$ then the resulting entropy is
constant while the energy is zero.\foot{ One could try to use an
alternative recipe by observing that since in the present case $p$ is
related to $k$ or $b$ and thus to $T$ (see \exxxx,\asy) one should
re-express everything in terms of $T$ before differentiating over it.
In that case \naii\ would take the form (see \asy) $ F' = - 4 \pi
e^{-2 \p_0} { T \ov 1- 8 \pi^2 \a' T^2}$ so that the entropy would be
$ S = 4 \pi e^{-2 \p_0} {1 + 8 \pi^2 \a' T^2 \ov 1- 8 \pi^2 \a'
T^2}$. The problem with this attempt is that for $\a'\to 0 $ one does
not reproduce the previous leading-order result for the energy (black
hole mass): the energy still vanishes in the $\a'\to 0$ limit. }

Another possibility is to follow \nap\ and modify the definition of the
 boundary term in the action \boni\ to make the subtraction of the
 infinite term automatic
\eqn\boni{
I'_\dM  = - 2  \int_0^{\b_0}  d \t  \  \del_x
 [( f-f_\infty)  e^{ -2 \pp}   ] |_{x= x_{bndry}} \ . } 
Compared to \boni\ we replaced $f$ by $f-f_\infty$, where $f_\infty$
is its asymptotic value, i.e. $(1-p)^{-1/2}$ in the case of \exann.
This gives, expanding  for  large $x$,
\eqn\gives{
 I'_\dM  = - 4 \pi e^{-2 \phi_0} { 1- 2 p \ov (1-p)^2}  e^{-2 b x} +
  O(e^{-4bx}) \ , }
 i.e. the subtracted value of the action and thus of the free energy
 is again zero for any $p$.  

\newsec{Free energy in  matrix model of 2-d black hole}

\subsec{\bf Noncritical $c_M=1$ string theory and   
 free energy from matrix model}

The matrix quantum mechanics (MQM) description of the 2-d string
theory on the black hole background \kkk\ appeared to be possible
due
to the duality, proposed in \FZZ,  between the two sigma models:
the
$({SL(2,C)\over SU(2)\times U(1)})_k$ ``Euclidean black
hole" coset model \witt\
%
%
%
%
and the Sine-Liouville
(SL) theory with the action (in this section $\a'$=1) 
\eqn\SL{{\cal S}_{SL}={1\over4\pi}\int d^2z \left[(\partial X)^2
+(\partial\vp)^2
+ Q\hat \CR\vp+\lambda e^{\gamma \vp}\ \cos\ R(X_L-X_R)\right]\ .  }
Here the scalar field $X$ (decomposed on shell as $X(z,\bar
z)=X_L(\bar z)+X_R(z)$) is compactified on a circle of radius $R$
and
$\vp$ is the ``Liouville field''. $\hat\CR=\sqrt g R^{(2)}$
 is a two dimensional
background curvature normalised so that $
{ 1 \ov 4 \pi}\int d^2 z   \hat\CR= 2-2h$, 
 where $h$ is the genus of the surface. 

The central charges of the two theories are to be equal, i.e.
$c={3k\over k-2}-1=1+6Q^2+1$ which gives $Q^2={1\over k-2}$.
To have the (1,1) conformal dimension
of the SL perturbation term
we have to put $\gamma =- Q^{-1} = -\sqrt{k-2}$ with 
 the
radius of compactification of $x$  being 
\eqn\RK{R=\sqrt k\ , }
i.e.  the same as the radius of the cigar far from the tip
(see \exann,\asy).

The MQM describes a theory  which is 
slightly different from the SL theory \SL:
\eqn\TWODS{{\cal S}_{c_M=1}={1\over4\pi}\int d^2 z
\left[(\partial X)^2 +(\partial\vp)^2
 + 2 \hat \CR\vp+\mu\vp e^{-2\vp}+\lambda e^{(R-2)\vp}\ 
\cos\ R(X_L-X_R)\right]\ . }
Here the central charge is always $c=26$, due to the choice of the
background charge $Q=2$, whereas the compactification radius $R$ of
$X$ is arbitrary within the interval $1<R<2$. This choice corresponds
to the central charge $c_M=1$ of the matter in the string theory.
Note that the coupling $\lambda$ plays the role of the fugacity of
vortices (windings)  with charges $\pm 1$ on the world sheet, similar
to the fugacity of vortices in the usual Sine-Gordon model.

However, the two theories \SL\ and \TWODS\ become the same for the
zero value of the ``cosmological constant" $\mu=0$ (or rather in the
limit $y={\mu \lambda^{2\over R-2}}\to 0$, see below) and the radius
$R=3/2$, corresponding to the level $k=9/4$ which is precisely the
string theory dilatonic black hole point in the coset description
\witt.

 In the matrix model approach of \kkk\ it was found that for the model
\TWODS\ the first-quantized string partition function $\CF(\mu,\l)$
(which, in space-time interpretation, is essentially the free energy,
up to a constant inverse temperature factor $\beta= 2\pi R$),
satisfies the Toda equation:
\eqn\TODA{
{1\over 4 }  \l^{-1}  \partial_\l \l\partial_\l
 \CF(\l,\mu) +\exp \left[ \CF(\l,\mu+i) + \CF(\l,\mu-i)-2 \CF(\l,\mu) \right]
=1 \ ,  }
with the boundary condition given  by the ${\mu}^{-1} \sim g_{s}$ expansion:
\eqn\FrenO{ 
\CF(\mu ,0)  = -  {R\ov 2}\mu^2 \log {\mu\over\Lambda} -
{1\over 24} \big(R +  R^{-1} \big)\log {\mu\over\Lambda}  +
R\sum_{h=2}^\infty \mu^{-2(h-1)} \tilde f^{(R)}_h+O(e^{-2\pi\mu})\
 .} 
Here $\tilde f^{(R)}_h$ are the known coefficients and $\Lambda$ is a
UV cutoff on the world sheet (we will drop the dependence on it
below).  The KPZ-DDK scaling following from the change of the
couplings induced by a shift of the zero mode of the Liouville field
$\vp$ imposes the constraints on the genus expansion of the free
energy of the theory \TWODS :
\eqn\FRENEX{ \CF(\l,\mu)=\sum_{h=0}^{\infty}\CF_h(\l, \mu)\ , }
where
\eqn\FRENH{\eqalign{
\CF_0(\l,\mu)& = {R\over 2-R} \mu^2 
\log\l + \l^{{4\over 2-R}} f_0(y)\ , \cr 
\CF_1(\l,\mu)& = -{R+R^{-1}\over 12(2-R)}\log\l+f_1(y)\ , \cr 
\CF_h(\l, \mu)& = (\l ^{2\over 2-R}) ^{2-2h}f_h(y)\ , \qquad \ \ \ 
 h\ge 2 \ ,  }}
with
 $$y \equiv {\mu \l^{-{2\over 2-R}}}\ .$$
The constants in front of the logarithmic terms are fixed by the boundary
conditions \FrenO\ and the KPZ-DDK scaling.

To isolate the genus 0 (2-sphere) contribution we may take $\mu$ very
large (by keeping $y$ fixed).  Then the equation \TODA\ becomes:
\eqn\TODAO{
{1\over 4 }  \l^{-1}  \partial_\l \l\partial_\l
 \CF(\l,\mu) +\exp \left[\partial_\mu^2\CF(\l,\mu)\right]=0.  }
 From here we can immediately find the free energy in the ``black
 hole'' limit $y \to 0$ (more precisely, for the Witten's black hole
 case we need also to set $R=3/2$ when \TWODS\ reduces to \SL ).  The
 relevant part of $\CF_0$ is given by $f_0(0)$ in \FRENH.  Plugging
 the logarithmic term from the first line of \FRENH\ into the
 r.h.s. of \TODAO\ we  find
\eqn\FREMO{  \F_0(\lambda,0)= -C(R)\ \lambda^{4\over 2-R} \ ,\ \ \
\ \ \ \ 
C(R)={1\over 4}(2-R)^2\ . }
%
The coefficient $C(R)$ can be, in principle, absorbed into the
dimensionful parameter $\lambda$.
\foot{Note that in principle we have
to change the sign of $\F_0$ to the opposite if we want to work in the
physically relevant canonical ensemble (fixed $N$ rather then fixed
$\mu$) within the MQM (see \kkk\ for the details).} However, once it
is chosen, it fixes all the subsequent terms in the genus expansion
\FRENH:
\eqn\FRENOO{ \CF(\lambda,0)= -C(R)\lambda^{4\over 2-R}-
{R+R^{-1}\over 48}\log\l^{4\over 2-R}+ \sum_{h=2}^\infty f^{(R)}_h 
\lambda^{4-4h\over 2-R}\ .}
Here $f^{(R)}_h=f_h(0)$ are universal functions of $R$ different from
$\tilde f$'s in \FrenO. They can be,  in principle,  determined from the
eq. \TODA\ (this has not yet been done).

Note that not only the scaling but also the sign
 of this matrix model
result for the free energy  \FREMO\
 is consistent with the expression \IFIN\ 
found from the ``dilaton subtraction'' approach in the effective 
field theory.

\subsec{\bf Comments on interpretation of the   matrix model result}


We conclude that the MQM approach to the bosonic string theory in the
black hole background gives a non-zero universal genus 0 string
partition function with the specific dependence on the
compactification radius $R$.

A thermodynamic interpretation of the black hole free energy depends
very much on  how one changes the temperature of the system.
At this stage we can take two different points of view on
thermodynamical interpretation of  the result \FRENOO :

(i) following the argument of \kkk\ we can assume that at least for
$R\to { 3\ov 2}$ the parameter $\lambda$ is a function of $R$,
adjusted in a way that
\eqn\FRENM{ \CF(\lambda)= 2\pi (R-{3\ov 2}) M -
{R+R^{-1}\over 24}\log M+ \sum_{h=2}^\infty \bar f^{(R)}_h M^{-h}
\ , }
where $M\sim [\lambda(R)]^{4\over 2-R}$ is an $R$-independent
parameter associated with the mass of the black hole. The 
coefficient $R-{3\ov 2}$ of the 2-sphere
term  is natural since it corresponds to the
result of the section 3  stating that the tree-level
string partition function (i.e. space-time effective action) is zero
at the Hawking temperature $T=T_H={1\over 3\pi}$ \ ($T={1\over 2\pi
R}$, see \asy). In that case the Hagedorn phase transition happens at
this temperature, the fluctuations of the energy are strong and we
have to reconsider the usual thermodynamics based on the Legendre
transform, taking into account the one loop logarithmic correction
(second term in the r.h.s of \FRENM ) and performing the integral over
these fluctuations. The details of this point of view on
thermodynamics of the 2-d black hole and some 
of its  physical consequences can
be found in \kkk .

(ii) we may try to adopt another point of view which seems more
appropriate  in the context of the 2-d string theory model \TWODS\ or its
MQM counterpart. Namely, we may take the $R$-dependence of the power
of $\l$ in \FREMO\ seriously (the $R$-dependence of the coefficient
$C(R)$ will not influence the following arguments) and view $\lambda$
as  $R$-independent free parameter of the theory. Let us consider
the partition function of an ensemble with fixed and equal numbers $
n$ of  vortices and anti-vortices of charges $\pm 1$, respectively,
which is represented  by the contour integral (around zero):
\eqn\EFR{   \exp\CF(n) = 
\oint {d\l\over 2\pi i \lambda^{2n+1}} \exp\CF_0(\lambda,0) \ , }
where  $\CF_0(\lambda,0)$ is given by  \FREMO .
$n$ shoud be even  for  the charge neutrality but it does not
matter for  large values of $n$ we shall consider here.  The saddle point
calculation (justified for large $n$) gives:
\eqn\FRn{   \CF(n) = (2-R)\ n\ \log {\Lambda\over\sqrt{n}}  +O(n)\  ,   }
where we restored the dependence on the cutoff $\Lambda$ resulting
from the non-universal terms in the MQM free energy.

It is natural to associate the $R$-dependent term in \FRn\ with
the mass term   $\beta M$
\eqn\MASS{ M={1\over 2\pi}\ n\ \log {\Lambda\over\sqrt{n}}\ ,}
while  the $R$-independent term - with the entropy $S$:
\eqn\ENTR{ S= 2n\ \log{\Lambda\over\sqrt{n}}  \ .}
We see that the free energy now vanishes not at the Hawking
temperature $R={3\ov 2} $, as in \FRENM , but rather at the
Kosterlitz-Thouless temperature $T_{KT} = { 1 \ov 2 \pi R_{KT} }, \ \
R_{KT}=2$.

These formulas predict that at the Hawking temperature corresponding
to $R={3 \ov 2} $ the free energy $F=-T\CF$ and the entropy are given by
\eqn\Fwq{ F= -{1\over 3} M \ , \ \ \ \ \ \ \ \ 
  S= 4\pi M \ . }
The finiteness and string coupling scaling of $F$ are consistent with
 the result
\IFIN\ following from the effective action approach 
(although they differ by sign in the ensemble with fixed $n$). The
expression for $S$ is different from $S= 3 \pi M$ implied by \ernt\ at
$T= T_H= { 1\ov 3\pi }$.

As a consequence of the FZZ conjecture \FZZ\ the entropy \ENTR\ counts
the number of states of the 2-d black hole.  This picture can be made
quite precise in the MQM language since we can explicitly write down
the Hamiltonian $\hat H_n$ of evolution in imaginary time $\beta=2\pi
R$ in the sector with fixed (conserved) number $n$ of vortex -
anti-vortex pairs (see Appendix).  The Gibbs partition function of MQM
is defined as
\eqn\HIGGS{  e^{\CF(n)} ={ \rm Tr}_n\ e^{-2\pi R \hat H_n}\ , }
where the trace is taken over the Hilbert space of MQM states
belonging to a reducible representation describing $n$ vortex -
anti-vortex pairs. Our estimate for  $\CF_n$ is given by the formula
\FRn\ extracted from the Toda equation. We see that the  non-singlet 
states have a large entropy $\sim M$ typical of a black hole. It would
be very instructive to derive it directly from the Hamiltonian given
in Appendix.

As we know from the collective field theory approach \DJP\ to the
singlet sector of the MQM,  such a Hamiltonian should produce an 
effective action for the tachyon scattering amplitudes. This would be
the most direct way to probe the background of this string theory.

Let us stress that we do not want to insist on this second point of
view as the only possible approach to thermodynamics of the 2-d black
hole. The above definition of the temperature is natural only in the
context of the model \TWODS\ or its MQM counterpart, since they have a
specific dependence of the string coupling on $R$.  When we vary the
radius $R$ (i.e. the inverse temperature) in this model we do not add
or remove any new matter fields from the system since the matter
central charge is always fixed to be $c_M=1$, while as in the coset
WZW model (and the Sine-Liouville model \SL) we change $c_M$ when
changing $R$, as is clear from  \RK . Also, the mass of the black hole in the
coset WZW model does not a priori depend on $R$.

To prefer one of
these two options we would  have to perform a truely
 microscopic calculation of thermodynamic properties of 
the system based on the MQM Hamiltonian.

\newsec{\bf Concluding remarks}

In this paper we tried to address some not yet well understood issues
concerning thermodynamic interpretation of the 2-d black hole in
string theory. The main purpose was to try to reconcile the old
result of the effective action approach to this problem
\gp\ (generalizing it to the case of the  exact black hole
  \dvv) and the new result following from the matrix model
approach \kkk .
 
We have shown that starting with the exact conformal 2-d \bh
background \exann,\eeee\ and subtracting the dilatonic vacuum
contribution from the free energy following the procedure of \gp\ one
finds the vanishing result.  At the same time, the matrix model
calculation of \kkk\ gave a non-zero result for the seemingly the same
quantity. One possibility to explain this discrepancy was suggested in
\kkk :  the non-vanishing matrix theory result for the tree-level
free energy should be interpreted as a non-universal contribution
which is to be omitted in comparing with space-time theory.

Hovever, the matrix theory free energy given by the solution of the Toda
equation is automatically finite and looks rather universal: the
non-universal IR divergent terms are already absent there.  This is
unlike the effective field theory expression containing the divergent
dilaton vacuum contribution \exrt.
The difference is obviously in how the subtraction of the divergent
vacuum contribution is done at the effective string field theory and
the matrix theory sides.  We suggested that the string partition
function in the matrix model is a different quantity from the one
given by the free energy \suub\ with the ``subtraction at the wall''
prescription of \gp\ --  the corresponding effective action 
  should instead 
be defined by the ``dilaton
subtraction'' prescription of section 3.2.  The advantage of the
latter recipe is that it is not related to any specific definition of
the temperature (which plays an important role in the approach of
\gp).\foot{It  was  conjectured  in 
\kkk\ that in 
 the matrix theory  analog  of the fermionic string
 (assuming it exists) 
 the expression for the sphere contribution to
the free energy 
should vanish. That
would be in agreement with the vanishing 
of the ``subtracted at the wall"  free energy
on the string theory side (note that the exact
\bh background in the supersymmetric case 
is given by \exn\ with $D-10 + 4 \a' b^2 =0, \ \ \a' b^2 = { 1 \ov k}
$). However, the  dilaton
 divergence subtraction of section 3.2 would still give
a non-vanishing result for the free energy in this case.
A resolution of this contradiction should await an actual construction 
of  fermionic string analog of MQM.
 }

We have also proposed in section 4.2 a ``microscopic'' estimate for
the entropy and the mass of the black hole based on fact that the Toda
equation solution in the matrix model can be interpreted as a
calculation of the ground state of the matrix Hamiltonian in the
relevant non-singlet representation of  the $U(N)$ symmetry of the
model.
 
It remains an open and interesting problem of how to derive the
boundary terms \boun\ in string theory effective action and thus the
\bh entropy (cf. \sug) directly from the string sigma model path
integral on 2-sphere.\foot{It may be useful to try to follow the
analogy with the particle theory path integral representation for the
logarithm of determinant of a Laplacian.  Indeed, the local
cutoff-dependent part of it is given by the Seeley coefficients
containing bulk and boundary terms.}  The local equations of motion
(beta function conditions of conformal invariance) do not ``know"
about the boundary terms in the effective action, so this is the case
which highlights the fundamental role of the string sigma model
partition function.

It is clear that we are still far from understanding of the general
picture explaining the thermodynamic behaviour of the 2-d black hole.
 Recent progress based on the matrix model formulation gives
hopes for a truely microscopic description of this interesting model.

\bigskip

{\bf Acknowledgements}

We are grateful to S. Alexandrov, A. Peet, J. Polchinski and
S. Shenker for useful discussions.  Part of this work was done while
we were participating in M-theory program at ITP, Santa Barbara
supported by the NSF grant PHY99-07949.  
The work of V.A.K. is
partially supported by the European Commission RTN programme 
 HPRN-CT-2000-00122.  
The work of A.A.T. is partially supported by the DOE grant
  DE-FG02-91ER40690 and also by 
   PPARC SPG  00613,  INTAS 991590
 and the  EC RTN   HPRN-CT-2000-00131 grants.   
%

\appendix{A} {Hamiltonian of the matrix model
in non-singlet representation}

In the matrix model approach of \refs{\kkk ,\BULKA}
the model \TWODS\ is described
by the MQM Hamiltonian
$$\hat H=-{1\over 2N}\Delta_X -{1\over 2}\tr X^2,$$ 
where $X$ is an $N\times N$ hermitean matrix coordinate and $\Delta_X$
is a Laplacian on the space of hermitean matrices.  The model has
the $SU(N)$ invariance and hence the states can be classified
according to representations of $SU(N)$. For an arbitrary
representation $r$ the Hamiltonian $\hat H_r$ can be expressed in
terms of the eigenvalues $z_1,\cdots,z_N$ of the matrix $X$ 
\eqn\HHAT{
H_r  = P_r \sum^N_{k=1} \big[ -{1 \over 2N} {\partial ^2 \over
\partial z_k^2} - {N\over 2}z_k^2 \big] + {1 \over 4N} \sum_{i < j}
{{\widehat
\tau }^r_{ij} {\widehat \tau }^r_{ij} \over (z_i - z_j)^2} \ .  }
$P_r$ is a projector onto the subspace of all zero weight vectors in
the space of representation (i.e. the kernel space of the generators
of the Cartan subalgebra) and ${\widehat \tau }^r_{ij}$ are the $N^2$
generators in this representation (see  \refs{\BULKA, \KAZ}
for details).

The (reducible) representation describing the dynamics of $n$ vortex
- anti-vortex pairs in the system corresponds to the character 
$$\chi_n(\Omega)=(\tr\Omega)^n(\tr\Omega^\dagger)^n\ ,$$ where
$\Omega\in Adj[SU(N)]$ (see \BULKA\ and \kkk\ for the explanation of
the connection between representations and vortex - anti-vortex
configurations). This representation (which we shall denote using
index $n$ instead of general label $r$) is just a direct product of
$n$ fundamental and $n$ anti-fundamental representations.  The
dimension of the representation is $N^{2n}$, the projector $P_n$ is
\eqn\PROJ{ 
P_n = (\underbrace{1\!\!1\times,\ldots,\times
1\!\!1}_{n},\underbrace{1\!\!1\times,\ldots,\times 1\!\!1}_{n}) \ , }
where $1\!\!1$ is the $N\times N$ unity matrix and the generators are
defined by the following action on an 
 arbitrary $N\times N$ matrix $A$
\eqn\GENS{ \eqalign{ \sum_{i,j=1}^N{\widehat \tau }^n_{ij}A_{ij}  
&=\sum_{m=1}^n
\big[ \ \underbrace{1\!\!1\times,\ldots,\times 1\!\!1}_{m-1}\times
A\times
 \underbrace{1\!\!1\times,\ldots,\times
 1\!\!1}_{n-m}\underbrace{1\!\!1\times,\ldots,\times
 1\!\!1}_{n}    \cr
&-\underbrace{1\!\!1\times,\ldots,\times 1\!\!1}_{n}
 \underbrace{1\!\!1\times,\ldots,\times 1\!\!1}_{m-1}\times
 A\times \underbrace{1\!\!1\times,\ldots,\times 1\!\!1}_{n-m} \
 \big]
. } }

\listrefs
\bye